\documentclass{article}
\usepackage{spconf}
\usepackage{amsmath}
\usepackage{graphicx}
\usepackage{multirow}
\usepackage{makecell}
\usepackage[flushleft]{threeparttable}
\usepackage{hyperref}
\usepackage{booktabs}
\usepackage{soul,color}
\usepackage[numbers]{natbib}
\usepackage{lipsum}
\usepackage{fancyhdr}
\usepackage[numbers]{natbib} 
\title{MarbleNet: Deep 1D Time-Channel Separable Convolutional Neural Network for Voice Activity Detection}
\name{Fei Jia , Somshubra Majumdar , Boris Ginsburg} 
 \address{NVIDIA, Santa Clara, USA\\
 \{fjia, smajumdar, bginsburg\}@nvidia.com
 }

\fancypagestyle{pageStyleOne}{%
    
    \fancyhf{}
    \fancyfoot[L]{\scriptsize Copyright 2021 IEEE. Published in ICASSP 2021 - 2021 IEEE International Conference on Acoustics, Speech and Signal Processing (ICASSP), scheduled for 6-11 June 2021 in Toronto, Ontario, Canada. Personal use of this material is permitted. However, permission to reprint/republish this material for advertising or promotional purposes or for creating new collective works for resale or redistribution to servers or lists, or to reuse any copyrighted component of this work in other works, must be obtained from the IEEE. Contact: Manager, Copyrights and Permissions / IEEE Service Center / 445 Hoes Lane / P.O. Box 1331 / Piscataway, NJ 08855-1331, USA. Telephone: + Intl. 908-562-3966.}
}

\begin{document}
\ninept
\maketitle
\begin{abstract}
We present \textit{MarbleNet}, an end-to-end neural network for Voice Activity Detection (VAD). 
% Robust voice detection in variable acoustic conditions is a vital first step for downstream speech processing applications. 
\textit{MarbleNet} is a deep residual network composed from blocks of 1D time-channel separable convolution, batch-normalization, ReLU and dropout layers.  
% \textit{MarbleNet} achieves state-of-the-art performance on AVA speech dataset, an evaluation benchmark for analysis of open domain media content on the web, while having significantly fewer parameters than similar models. 
When compared to a state-of-the-art VAD model, \textit{MarbleNet} is able to achieve similar performance with roughly 1/10-th the parameter cost.
We further conduct extensive ablation studies on different training methods and choices of parameters in order to study the robustness of \textit{MarbleNet} in real-world VAD tasks. 

\end{abstract}
\section{Introduction}

\label{sec:intro}
Voice Activity Detection (VAD), also known as \textit{speech activity detection} or \textit{speech detection}, is a binary classification task of inferring which segments of input audio contain speech versus which segments are background noise.
It is an essential first step for a variety of downstream speech-based applications such as automatic speech recognition and speaker diarization.
A typical VAD system is a frame-level classifier using acoustic features to determine whether the frame (usually an audio segment of 10ms duration) belongs to a speech or non-speech class.

Apart from feature-based~\citep{graf2015features, ramirez2004efficient, moattar2009simple} and statistical modeling approaches~\citep{statistical, sohn1999statistical}, recent research effort has been devoted to finding efficient deep-learning-based VAD model architectures.
Notable examples include Recurrent Neural Networks (RNN)~\cite{rnn, lstm_movie, EndToEndDAVAD2020}, Convolutional Neural Networks (CNN) ~\cite{ava, temporal, robust, cnn_mismatch}, and Convolutional Long Short-Term Memory (LSTM) Deep Neural Networks (CLDNN)~\cite{cldnn}, which conduct frequency modeling with CNN and temporal modeling with LSTM. 
LSTM is a popular choice for sequential modeling of VAD tasks~\citep{cldnn, rnn}.
It has been observed that LSTMs suffer from state saturation problems when the utterance is long.
To address this issue, \citet{temporal} proposed a stateless dilated CNN (with 400K parameters) for temporal modeling.
\citet{cnn_mismatch} found severe performance degradation when the VAD models were evaluated on unseen audio recorded from radio channels.
They also demonstrated that CNNs were useful acoustic models in novel channel scenarios and able to adapt well with limited amounts of  data.
\citet{robust} compared different LSTM and CNN models with more challenging movie data which contained post-production stage and atypical speech such as electronically modified speech samples.
They proposed a Convolutional Neural Network-Time Distributed (CNN-TD) model (with 740K parameters) that outperformed existing models including Bi-LSTM (with 300K parameters), CLDNN (with 1M parameters) and ResNet 960 (with 30M parameters)~\citep{ava} on the benchmark evaluation dataset \textit{AVA-speech}.
Furthermore, acoustic models using 1D CNNs have shown great potential in automatic speech recognition~\cite{kriman2019quartznet,li2019jasper,Wav2LetterV2} and speech command detection~\cite{matchboxnet} tasks.

Built on top of previous successful applications of  1D CNNs to speech processing  tasks,
we introduce \textit{MarbleNet}, a compact end-to-end neural network for VAD inspired by the QuartzNet architecture~\citep{kriman2019quartznet} and the MatchboxNet model~\citep{matchboxnet}. 
\textit{MarbleNet} is constructed with a stack of blocks with residual connections~\cite{he2015}.
Each block is composed from 1D time-channel separable convolutions, batch normalization, ReLU, and dropout layers.
Those 1D time-channel separable convolutions are similar to 2D depth-wise separable convolutions~\citep{chollet2017xception,kaiser2017depthwise}.
The usage of 1D time-channel separable convolutions significantly reduces the number of parameters in the model. \textit{MarbleNet}'s size is 1/10-th of a state-of-the-art VAD model CNN-TD~\cite{robust} in terms of number of parameters, or 1/35-th in terms of MFLOPs, enabling more efficient model inference and storage.
With increasingly widespread usage of speech-based applications across mobile and wearable devices (e.g., virtual assistant), building lightweight VAD models carries important implications for real-world applications in memory and compute constrained deployment scenarios.

Specifically, this paper makes the following contributions:
\begin{enumerate}
  \item We propose MarbleNet: a novel end-to-end neural model for VAD based on 1D time-channel separable convolutions.
  \item  MarbleNet achieves state-of-the-art performance on the AVA speech dataset~\citep{ava}, and it has 10x fewer  parameters compared to the CNN-TD  model~\citep{robust}. 
  \item We perform ablation studies for model and training parameters including input features, noise augmentation, and overlapped predictions.
  \item We open source the model, training and inference pipelines together with pre-trained checkpoints for easy reproduction of presented results.
\end{enumerate}

\section{Model Architecture}
\label{sec:architecture}

\begin{figure}[t]
  \centering
  \includegraphics[width=\linewidth]{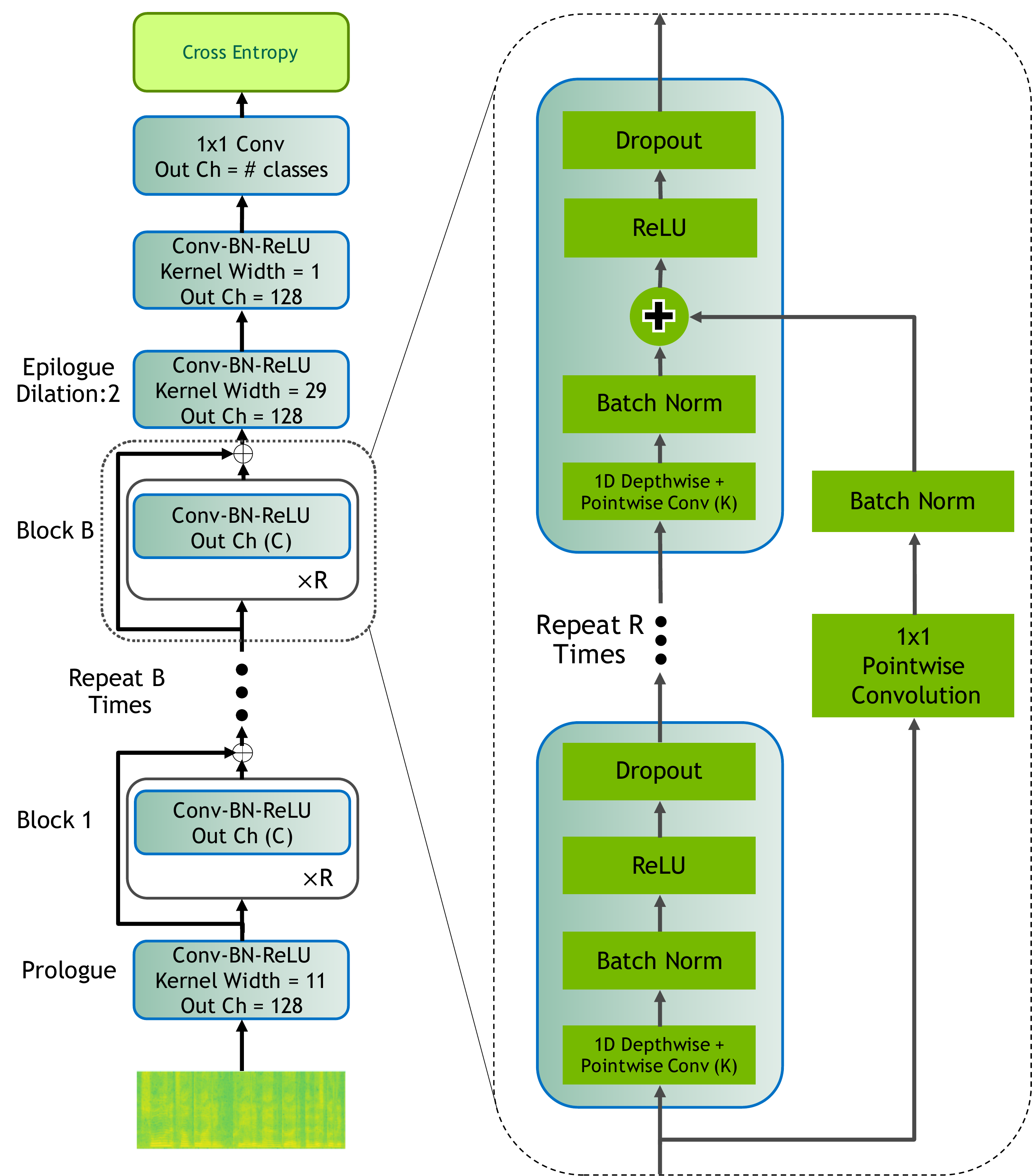}
  \caption{MarbleNet $B$x$R$x$C$ model: $B$ - number of blocks, \quad $R$ - number of sub-blocks, $C$ - the number of channels.}
  \label{fig:quartznet_arch}
\end{figure}

MarbleNet-$B$x$R$x$C$ is based on the QuartzNet architecture. It includes $B$ residual blocks each with $R$ sub-blocks. All sub-blocks within each block have the same $C$ output channels (see Fig.~\ref{fig:quartznet_arch}).
A basic sub-block consists of a 1D-time-channel separable convolution, 1x1 pointwise convolutions, batch norm, ReLU, and dropout. The 1D time-channel separable convolution has $C$ filters with a kernel of the size $K$.
All models have four additional sub-blocks: one prologue layer -- `Conv1' before the first block, and three epilogue sub-blocks  (`Conv2', `Conv3', and `Conv4') before the final soft-max layer.
For example, the complete architecture for \textit{MarbleNet}-3x2x64 (B=3 blocks, R=2 sub-blocks per block, C=64 channels) is shown in the Table~\ref{tab:QuartzNetParams}.

% {\renewcommand{\arraystretch}{1.1}
\begin{table}[ht]
\caption{\textit{MarbleNet}-3x2x64 model has B=3 blocks, each block has R=2 time-channel separable convolutional sub-blocks with C=64 channels, plus 4 additional sub-blocks: prologue - Conv1, and epilogue -  Conv2, Conv3, Conv4).}
\label{tab:QuartzNetParams}
\centering
\scalebox{0.8}{
\begin{tabular}{c c c c c} 
 \toprule
  \textbf{Block} &
  \textbf{\# Blocks} &
  \textbf{\thead{\# Sub\\Blocks}}  &
  \textbf{\thead{\# Output\\Channels}} &
  \textbf{Kernel} \\
 \midrule
 Conv1&1 & 1 & 128 & 11 \\
 B1   &1 & 2 & 64 & 13 \\
 B2   &1 & 2 & 64 & 15 \\
 B3   &1 & 2 & 64 & 17 \\
 Conv2 & 1 & 1 & 128 & 29, \textit{dilation=2}\\
 Conv3 & 1 & 1 & 128 & 1 \\
 Conv4 &1 &  1 & \# classes & 1 \\
 Soft-max &\\
 Cross-entropy &\\
 \bottomrule
\end{tabular}
}
\end{table}
% }

\section{Experiments}
\label{sec:experiments}
%\subsection{Experiment Setup}
\subsection{Training Data}

Generally, noise-robust VAD systems are developed using audio from clean speech datasets augmented with different types of noise~\citep{ding2020personal, robust, EndToEndDAVAD2020}.
\citet{robust} compiled the Subtitle-Aligned Movie (SAM) corpus, a dataset based on 117 hours of movie audio.
%(comprised of approximately 20\% speech content).
% SAM contains coarsely labelled speech and non-speech regions and has around 63,500 speech segments, each of 1.28s in duration. 
% The authors used true positive rate (TPR) for a given false positive rate (FPR) equals to 0.315 as a evaluation metric.
% And they conclude the TPR, of 0.77 for FPR=0.315 on AVA-speech, obtained by their models trained on the noise-augmented dataset (audiobook data from MUSAN~\citep{musan}) is significantly lower than that of the models trained on SAM corpus. 
However, the SAM  dataset is constructed from movies, some of which are restricted from commercial usage. 
Therefore, we use Google Speech Commands Dataset V2~\cite{warden2018speech} as speech data. This dataset contains 105,000 utterances, each approximately 1s long, belonging to one of 35 classes of common words like ``Yes" and ``Go". 
Around 2,700 variable-length samples of 35 background categories such as ``traffic noise", ``inside, small room" from freesound.org~\cite{freesound} served as non-speech data. Each sample was between 0.63s to 100s. The speech commands dataset and Freesound dataset are referred to as the \textit{SCF} dataset hereafter.
We split SCF into train, validation and test sets using an 8:1:1 ratio.

Following~\citet{robust}, the samples were converted to segments of length 0.63s, so that the input 64 dimensional mel-filterbank features calculated from 25ms
windows with a 10ms overlap would be of shape 64x64 when provided to CNN models, for fair comparison with~\citet{robust}. For speech samples in the train and validation sets, we used the central segment (0.2s-0.83s) of each sample. Speech segments in the test set and non-speech segments were generated with stride length 0.15s. 
We re-balanced SCF speech and non-speech classes to have the same number of segments in each of the classes. In total, 160,994, 20,018 and 60,906 segments were used for training, validation and testing respectively.

\subsection{Training Methodology}
\label{ssec:training_methodology}

We pre-processed the audio segments with 64 MFCC features for training \textit{MarbleNet}.
Then, with a probability of 80\%, the input was augmented with time shift perturbations in the range of $T=[-5, 5]$ ms and white noise of magnitude $[-90, -46]$ dB. 
Additionally, we applied SpecAugment~\citep{park2019} with 2 continuous time masks of size $[0, 25]$ time steps, and 2 continuous frequency masks of size $[0, 15]$ frequency bands. 
We also used SpecCutout~\citep{devries2017specutout} with 5 rectangular masks in the time and frequency dimensions.

All models were trained with the SGD optimizer with momentum \citep{momentum}, using $\text{momentum}=0.9$ and $\text{weight decay}=0.001$. 
We utilized the Warmup-Hold-Decay learning rate schedule \citep{he2019bag} with a warm-up ratio of 5\%, a hold ratio of 45\%, and a polynomial (2nd order) decay for the remaining 50\% of the schedule. 
A maximum learning rate of 0.01 and a minimum learning rate of 0.001 were used.
We trained all models for 150 epochs on 2 V100 GPUs with a batch size of 128 per GPU. 
The model was implemented and trained  with  NeMo, an open-source toolkit for  Conversational AI~\cite{nemo2019}.\footnote{ \url{https://github.com/NVIDIA/NeMo}.}

% {\renewcommand{\arraystretch}{1.1}
\begin{table*}[ht]
\caption{Model performance on the AVA-speech dataset~\citep{ava}.
We report the true positive rate (TPR) when the false positive rate (FPR) equals 0.315 to compare results with~\citet{robust}, and report AUROC for the ``All" category in AVA-speech to demonstrate the overall TPR-FPR trade-off.
CNN-TD follows the reported methodology in~\citet{robust} and is obtained after a hyperparameter search using the SCF validation set. 
We generate predictions by overlapping input segments with 87.5\% overlap and adopting a median smoothing filter.
Each result is averaged over 5 trials (95\% Confidence Interval). Since we observed a large variance in the scores for the CNN-TD model, we report the results averaged over 10 trials.}
\label{tab:results_mbn}
\centering
\vspace{0.1in}
\scalebox{0.8}{
\begin{tabular}{ l| c |c c c c | c} 
 \toprule
%  \multicolumn{2}{c}{} & \multicolumn{5}{c}{\textbf{AVA speech}}\\
%  \midrule
 \textbf {Model} & \textbf{\thead{\# Parameters}} &\multicolumn{4}{c}{\textbf{TPR for FPR = 0.315}}  & \textbf{ AUROC} \\
 & \textbf{\thead{(K)}} &  \textbf{Clean} &\textbf{+Noise} &\textbf{+Music} &\textbf{All} &\textbf{All}\\
  \midrule
CNN-TD & 738  & 0.911$\pm$0.063  &  0.795$\pm$0.056 &  0.797$\pm$0.048 & 0.827$\pm$0.055 & 0.821$\pm$0.055 \\
CNN-TD + 87.5$\%$ median   & 738  & 0.935$\pm$0.057  & \textbf{0.824$\pm$0.051} & 0.824$\pm$0.043  &0.855$\pm$0.050 &0.841$\pm$0.050 \\
\midrule
\textit{MarbleNet}-3x2x64                   & 88  & 0.924$\pm$0.005  & 0.815$\pm$0.014 & 0.822$\pm$0.017  & 0.847$\pm$0.012 & 0.850$\pm$0.009 \\ 
\textit{MarbleNet}-3x2x64 + 87.5$\%$ median & \textbf{88}  & \textbf{0.942 $\pm$0.008}  & 0.821$\pm$0.022  &  \textbf{0.834$\pm$0.016}  &  \textbf{0.858$\pm$0.016} &\textbf{0.858$\pm$0.011}   \\
\bottomrule
\end{tabular}
}
\end{table*}
% }
\subsection{Evaluation Method}
\label{ssec:evaluation}
We use the CNN-TD model proposed by \citet{robust} as our baseline.
In order to compare our model with CNN-TD and examine the ability of the model to generalize to samples outside of the training domain, we evaluate our models on the AVA-speech dataset~\citep{ava}.
% \textit{MarbleNet} achieves state-of-the-art performance on AVA speech dataset, an evaluation benchmark for analysis of open domain media content on the web, while having significantly fewer parameters than similar models. 
AVA-speech is a public, multi-language, densely labelled dataset of YouTube videos with one non-speech class and three speech conditions: clean speech (``clean''), speech with noise (``+noise''), and speech with music (``+music''). 
Each instance in the AVA-speech dataset is a 15-minute movie clip. It has a natural mix of background noise conditions, as opposed to synthetic (white/pink) noise.
The dataset only contains labels for video segments, and in order to evaluate models, we still need to access the corresponding videos on YouTube.
We use 122 out of 160 labelled movies that are still publicly available on YouTube at the time of the experiment as our AVA-speech evaluation dataset, totaling 30 hours of playback time.
For each speech condition, we calculate the true positive rate (TPR) at the frame level contrasted with non-speech as the negative class. Moreover, we combine all the frames for the 3
speech conditions into a single positive class (``All").

We follow \citet{robust} and report the true positive rate (TPR) when the false positive rate (FPR) equals 0.315.
Since the TPR for FPR = 0.315 is only a single point on the receiver operating characteristic (ROC) curve, we also report the area under ROC (AUROC) representing the overall TPR-FPR trade-off.

To eliminate the potential dataset quality difference, instead of using the saved checkpoint trained on a different dataset, we retrained and tuned the CNN-TD model on our SCF train set.
We followed the exact training procedure as the authors describe in their paper.
During model training, we used a batch-size of 64, and a binary cross-entropy loss function.
Mel spectrogram was used as the input feature.
We trained the network for 5 epochs and performed early stopping if validation loss did not decrease by 1e-3 for 3 consecutive epochs.
The Adam~\cite{kingma2014adam} optimizer was used with learning rate 1e-4.
We note that we also experimented extensively with other training procedures, input features, and hyper-parameter choices, and eventually determined that the default setup described above produced the best performing model for the CNN-TD baseline.
CNN-TD experiment results reported below are produced with the above setup.

% http://www.ugr.es/~segura/pdfdocs/specom04.pdf
% The 8 kHz input signal was decomposed into overlapping frames with a 10-ms
% window shift.  speech and noise distributions are better separated when increasing the order of the long-term window. The speech classification error is approximately reduced by half from 22\% to 9\% when the order of the VAD is increased from 0 to 6 frames. 

During inference, we performed frame-level prediction by two approaches:
1) shift the window by 10ms to generate the frame and use the prediction of the window to represent the label for the frame;
2) generate predictions with overlapping input segments. Then a smoothing filter is applied to decide the label for a frame spanned by multiple segments. We examined two common smoothing filters: majority vote (median)~\citep{robust} and average (mean)~\citep{EndToEndDAVAD2020}. We also experimented with different amounts of overlap ranging from 12.5\% to 87.5\% to understand the effect of this parameter for frame-level performance. The results are shown in Table~\ref{tab:tab:overlapped_pred_table}.

\subsection{Results}
Table~\ref{tab:results_mbn} shows the evaluation results on the AVA-speech dataset.
We repeated each trial 5 times for \textit{MarbleNet}, and 10 times for CNN-TD as we observed greater variance in the performance of this model.
\textit{MarbleNet} results display a  smaller confidence interval, suggesting \textit{MarbleNet} achieves stable training over many runs.

\textit{MarbleNet}-3x2x64 achieves similar performance as CNN-TD but has roughly 1/10-th the number of parameters.
This result demonstrates that even when trained on a relatively simple training dataset (SCF train set), \textit{MarbleNet}-3x2x64 can still obtain good test performance for the more challenging  AVA-speech dataset without additional fine-tuning.

\subsection{Ablation Study}
We conducted a series of ablation studies to investigate the impact of input features, inference methods, and noise augmentation on model performance.
\subsubsection{MFCC vs Mel spectrogram}
\begin{figure*}[ht]
  \centering
  \includegraphics[width=1.0\linewidth]{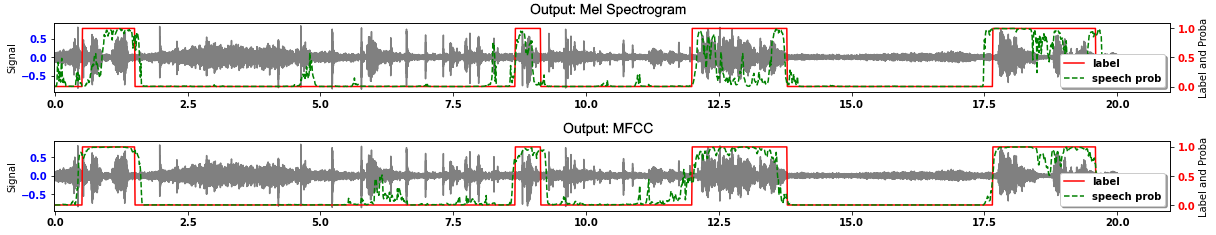}
  \caption{Output plot of Mel spectrogram and MFCC. Example is from AVA-speech. The first two segments are speech with noise, the last two are speech with music.}
  \label{fig:feature}
\end{figure*}

We first compare the two choices of potential preprocessing and feature extraction methods for audio signals: Mel spectrogram and MFCC.
They are examined with a \textit{MarbleNet}-3x2x64 model.
As shown in Table~\ref{tab:mfcc_vs_mel}, MFCC outperforms Mel spectrogram on AVA-speech.
When FPR equals 0.315, MFCC increases TPR by 0.030 on ``clean'' speech and has even greater improvements of 0.078 and 0.079 on two ``noisy'' speech classes respectively.
These results align with the fact that the compressible lower-order cepstral coefficients of MFCC represent simple aspects of the spectral envelope.
Coefficients in MFCC are less correlated and contain most of the information about the overall spectral shape.
In contrast, high-order coefficients in Mel spectrogram are closer to noise and of less importance.
We show an example in Figure~\ref{fig:feature}, and it is visually apparent that MFCC (bottom panel) produces better inference than Mel spectrogram (top panel).

\begin{table}[!h]
\caption{Model performance of  \textit{MarbleNet}-3x2x64 under the same training methodology but with two different input features, MFCC and Mel spectrogram, on AVA-speech dataset.
Each result is averaged over 5 trials (95\% Confidence Interval).}
\label{tab:mfcc_vs_mel}
\centering
\vspace{0.1in}
\scalebox{0.75}{
\begin{tabular}{c | c c c c |c} 
 \toprule
 \textbf {Feature} &\multicolumn{4}{c}{\textbf{TPR for FPR=0.315}} & \textbf{AUROC}  \\
& \textbf{Clean} &\textbf{+Noise} &\textbf{+Music} &\textbf{All} &\textbf{All} \\
  \midrule
 MFCC & 0.92$\pm$0.01  & 0.82$\pm$0.01 & 0.82$\pm$0.02  & 0.85$\pm$0.01 & 0.85$\pm$0.01 \\ 
 Mel spectrogram & 0.89$\pm$0.02   & 0.74$\pm$0.02  & 0.74$\pm$0.03  &0.77$\pm$0.04 & 0.80$\pm$0.01 \\
 \bottomrule
\end{tabular}
}
\end{table}

\subsubsection{Segment length}
People speak at different speeds and their pauses might be shorter than our segment length 0.63s.
If we want to detect short pauses or a speaker change within a short time, a smaller segment length could help. 
Since our model is fully convolutional, instead of re-training on a smaller segment length, we can simply feed in smaller segments as input data for inference. 
Nevertheless, because AVA-speech's ground truth labels do not account for very short pauses, using shorter input segments does not significantly affect the model's performance.

\begin{figure}[ht]
  \centering
  \includegraphics[width=1\linewidth]{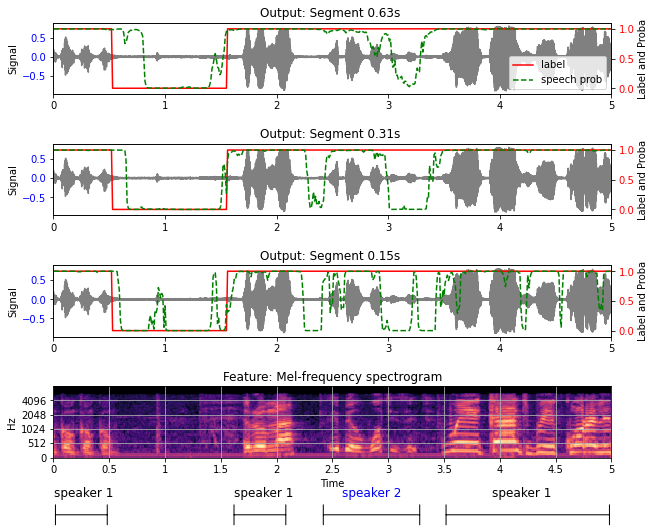}
  \caption{Output plot of different segment lengths. Example is from AVA-speech.}
  \label{fig:seg}
\end{figure}

Figure~\ref{fig:seg} shows such an example of varying input segment lengths.
Shorter segment inputs indeed allow us to detect more pauses (green lines), however the ground truth labels (red lines) mark the entire latter part of the example as ``speech''.

\subsubsection{Overlapping Predictions}
Median and mean smoothing filters are generally used when generating frame level predictions~\citep{smartapp, EndToEndDAVAD2020}. In Table~\ref{tab:tab:overlapped_pred_table}, we show the results of these two filters using three different amounts of overlap: 12.5\%, 50.0\%, and 87.5\%.
We can see that model performance increases as the overlap increases. There is no significant difference between these two smoothing filters for the 87.5\% overlap case.
 
{
\begin{table}[!h]
\caption{Model performance of \textit{MarbleNet}-3x2x64 using mean and median filters with 12.5\%, 50\%, and 87.5\% overlap during inference on AVA-speech.}
\vspace{0.1in}
\label{tab:tab:overlapped_pred_table}
\centering
\scalebox{0.8}{
\begin{tabular}{  l |c c c c | c} 
 \toprule
 \textbf {Smoothing filter}   &\multicolumn{4}{c}{\textbf{TPR for FPR=0.315}}  &\textbf{AUROC} \\
  &  \textbf{Clean} &\textbf{+Noise} &\textbf{+Music} &\textbf{All} &\textbf{All}\\
\midrule
Median  0.125 & 0.915 & 0.822 &  0.831 &  0.850 & 0.851 \\ 
Median  0.500 & 0.933 & 0.829 &  0.837 &  0.860 & 0.861 \\ 
Median 0.875 & 0.943 & 0.852 &  0.857 & 0.879 & 0.876 \\ 
\midrule
Mean 0.875 & 0.938 & 0.832 & 0.840 & 0.863 &  0.866 \\
\bottomrule
\end{tabular}
}
\end{table}
}

\subsubsection{Noise Augmented Training}
\label{sec:noise_augmentation}
We compare \textit{MarbleNet}-3x2x64 trained with the basic augmentation described in Section~\ref{ssec:training_methodology}, and without augmentation.  Table~\ref{tab:Noise_augmented_training} indicates that synthetic noise augmentation is helpful to improve the performance.

{
\begin{table}[!h]
\caption{\textit{MarbleNet}-3x2x64 trained with and without noise augmentation.
Each result is averaged over 5 trials (95\% Confidence Interval).}
\vspace{0.1in}
\label{tab:Noise_augmented_training}
\centering
\scalebox{0.8}{
\begin{tabular}{  l |c c c c | c} 
 \toprule
 \textbf {Augmen-}   &\multicolumn{4}{c}{\textbf{TPR for FPR=0.315}}  &\textbf{AUROC} \\
  \textbf {tation} &  \textbf{Clean} &\textbf{+Noise} &\textbf{+Music} &\textbf{All} &\textbf{All}\\
\midrule
Basic & 0.92$\pm$0.01  & 0.82$\pm$0.01 & 0.82$\pm$0.02  & 0.85$\pm$0.01 & 0.85$\pm$0.01 \\
None  & 0.90$\pm$0.01 & 0.76$\pm0.01$ & 0.79$\pm$0.01 &  0.80$\pm$0.01 & 0.83$\pm$0.00 \\ 
\bottomrule
\end{tabular}
}
\end{table}
}

% \section{Discussion}
% \label{sec:discussion}
\section{Conclusions}
\label{sec:conclusion}
In this paper, we present \textit{MarbleNet}, a computationally efficient, state-of-the-art deep Voice Activity Detection model.
We evaluate \textit{MarbleNet} on a diverse, multi-language, natural-sounding AVA-speech movie dataset.
We demonstrate that \textit{MarbleNet} is able to achieve state-of-the-art performance with 1/10th the parameters compared to the baseline model, which can make VAD more accessible under computationally constrained environments such as mobile and wearable devices.
We conduct an extensive series of ablation studies to inspect the model's behavior and serve as a guideline for future researchers and practitioners.
Finally, by building the efficient yet lightweight \textit{MarbleNet} with the open-source NeMo framework,
we hope our work can facilitate wider adaptation of modern VAD technologies and democratization of AI.

\vskip 0.2in
\noindent \textbf{Acknowledgement}
We would like to thank the NVIDIA AI Applications team for the help and valuable feedback.

% \vfill\pagebreak

\label{sec:refs}

\bibliographystyle{IEEEtranN}
% \bibliography{strings,refs}
\bibliography{vadbib}

% Generated by IEEEtranN.bst, version: 1.14 (2015/08/26)
\begin{thebibliography}{30}
\providecommand{\natexlab}[1]{#1}
\providecommand{\url}[1]{#1}
\csname url@samestyle\endcsname
\providecommand{\newblock}{\relax}
\providecommand{\bibinfo}[2]{#2}
\providecommand{\BIBentrySTDinterwordspacing}{\spaceskip=0pt\relax}
\providecommand{\BIBentryALTinterwordstretchfactor}{4}
\providecommand{\BIBentryALTinterwordspacing}{\spaceskip=\fontdimen2\font plus
\BIBentryALTinterwordstretchfactor\fontdimen3\font minus
  \fontdimen4\font\relax}
\providecommand{\BIBforeignlanguage}[2]{{%
\expandafter\ifx\csname l@#1\endcsname\relax
\typeout{** WARNING: IEEEtranN.bst: No hyphenation pattern has been}%
\typeout{** loaded for the language `#1'. Using the pattern for}%
\typeout{** the default language instead.}%
\else
\language=\csname l@#1\endcsname
\fi
#2}}
\providecommand{\BIBdecl}{\relax}
\BIBdecl

\bibitem[Graf et~al.(2015)Graf, Herbig, Buck, and Schmidt]{graf2015features}
S.~Graf, T.~Herbig, M.~Buck, and G.~Schmidt, ``Features for voice activity
  detection: a comparative analysis,'' \emph{EURASIP Journal on Advances in
  Signal Processing}, vol. 2015, no.~1, pp. 1--15, 2015.

\bibitem[Ram{\i}rez et~al.(2004)Ram{\i}rez, Segura, Ben{\i}tez, De~La~Torre,
  and Rubio]{ramirez2004efficient}
J.~Ram{\i}rez, J.~C. Segura, C.~Ben{\i}tez, A.~De~La~Torre, and A.~Rubio,
  ``Efficient voice activity detection algorithms using long-term speech
  information,'' \emph{Speech communication}, vol.~42, no. 3-4, pp. 271--287,
  2004.

\bibitem[Moattar and Homayounpour(2009)]{moattar2009simple}
M.~H. Moattar and M.~M. Homayounpour, ``A simple but efficient real-time voice
  activity detection algorithm,'' in \emph{2009 17th European Signal Processing
  Conference}.\hskip 1em plus 0.5em minus 0.4em\relax IEEE, 2009, pp.
  2549--2553.

\bibitem[Ramirez et~al.(2007)Ramirez, G{\'o}rriz, and Segura]{statistical}
J.~Ramirez, J.~G{\'o}rriz, and J.~Segura, ``Voice activity detection.
  fundamentals and speech recognition system robustness,'' \emph{Robust speech
  recognition and understanding}, vol.~6, no.~9, pp. 1--22, 2007.

\bibitem[Sohn et~al.(1999)Sohn, Kim, and Sung]{sohn1999statistical}
J.~Sohn, N.~Kim, and W.~Sung, ``A statistical model-based voice activity
  detection,'' \emph{IEEE signal processing letters}, vol.~6, no.~1, pp. 1--3,
  1999.

\bibitem[{Gelly} and {Gauvain}(2018)]{rnn}
G.~{Gelly} and J.~{Gauvain}, ``Optimization of rnn-based speech activity
  detection,'' \emph{IEEE/ACM Transactions on Audio, Speech, and Language
  Processing}, vol.~26, no.~3, pp. 646--656, 2018.

\bibitem[{Eyben} et~al.(2013){Eyben}, {Weninger}, {Squartini}, and
  {Schuller}]{lstm_movie}
F.~{Eyben}, F.~{Weninger}, S.~{Squartini}, and B.~{Schuller}, ``Real-life voice
  activity detection with {LSTM} recurrent neural networks and an application
  to hollywood movies,'' in \emph{ICASSP}, 2013.

\bibitem[Lavechin et~al.(2020)Lavechin, Bousbib, Bredin, Dupoux, Cristia, Gill,
  and Garcia-Perera]{EndToEndDAVAD2020}
M.~Lavechin, R.~Bousbib, H.~Bredin, E.~Dupoux, A.~Cristia, M.-P. Gill, and
  L.~P. Garcia-Perera, ``End-to-end domain-adversarial voice activity
  detection,'' in \emph{Interspeech 2020}, 2020.

\bibitem[Chaudhuri et~al.(2018)Chaudhuri, Roth, Ellis, Gallagher, Kaver,
  Marvin, Pantofaru, Reale, {Guarino Reid}, Wilson, and Xi]{ava}
S.~Chaudhuri, J.~Roth, D.~P.~W. Ellis, A.~Gallagher, L.~Kaver, R.~Marvin,
  C.~Pantofaru, N.~Reale, L.~{Guarino Reid}, K.~Wilson, and Z.~Xi,
  ``Ava-speech: A densely labeled dataset of speech activity in movies,'' in
  \emph{Proc. Interspeech 2018}, 2018, pp. 1239--1243.

\bibitem[{Chang} et~al.(2018){Chang}, {Li}, {Simko}, {Sainath}, {Tripathi},
  {van den Oord}, and {Vinyals}]{temporal}
S.~{Chang}, B.~{Li}, G.~{Simko}, T.~N. {Sainath}, A.~{Tripathi}, A.~{van den
  Oord}, and O.~{Vinyals}, ``Temporal modeling using dilated convolution and
  gating for voice-activity-detection,'' in \emph{ICASSP}, 2018.

\bibitem[{Hebbar} et~al.(2019){Hebbar}, {Somandepalli}, and
  {Narayanan}]{robust}
R.~{Hebbar}, K.~{Somandepalli}, and S.~{Narayanan}, ``Robust speech activity
  detection in movie audio: Data resources and experimental evaluation,'' in
  \emph{ICASSP}, 2019.

\bibitem[{Thomas} et~al.(2014){Thomas}, {Ganapathy}, {Saon}, and
  {Soltau}]{cnn_mismatch}
S.~{Thomas}, S.~{Ganapathy}, G.~{Saon}, and H.~{Soltau}, ``Analyzing
  convolutional neural networks for speech activity detection in mismatched
  acoustic conditions,'' in \emph{ICASSP}, 2014.

\bibitem[Zazo-Candil et~al.(2016)Zazo-Candil, Sainath, Simko, and
  Parada]{cldnn}
R.~Zazo-Candil, T.~Sainath, G.~Simko, and C.~Parada, ``Feature learning with
  raw-waveform {CLDNN}s for voice activity detection,'' in \emph{INTERSPEECH},
  2016.

\bibitem[Kriman et~al.(2020)Kriman, Beliaev, Ginsburg, Huang, Kuchaiev,
  Lavrukhin, Leary, Li, and Zhang]{kriman2019quartznet}
S.~Kriman, S.~Beliaev, B.~Ginsburg, J.~Huang, O.~Kuchaiev, V.~Lavrukhin,
  R.~Leary, J.~Li, and Y.~Zhang, ``Quartznet: Deep automatic speech recognition
  with 1d time-channel separable convolutions,'' in \emph{ICASSP 2020-2020 IEEE
  International Conference on Acoustics, Speech and Signal Processing
  (ICASSP)}.\hskip 1em plus 0.5em minus 0.4em\relax IEEE, 2020, pp. 6124--6128.

\bibitem[Li et~al.(2019)Li, Lavrukhin, Ginsburg, Leary, Kuchaiev, Cohen,
  Nguyen, and Gadde]{li2019jasper}
J.~Li, V.~Lavrukhin, B.~Ginsburg, R.~Leary, O.~Kuchaiev, J.~M. Cohen,
  H.~Nguyen, and R.~T. Gadde, ``Jasper: An end-to-end convolutional neural
  acoustic model,'' \emph{Proc. Interspeech 2019}, pp. 71--75, 2019.

\bibitem[Liptchinsky et~al.(2017)Liptchinsky, Synnaeve, and
  Collobert]{Wav2LetterV2}
V.~Liptchinsky, G.~Synnaeve, and R.~Collobert, ``Letter-based speech
  recognition with gated convnets,'' \emph{arxiv:1712.09444}, 2017.

\bibitem[Majumdar and Ginsburg(2020)]{matchboxnet}
S.~Majumdar and B.~Ginsburg, ``{MatchboxNet}: $1d$ time-channel separable
  convolutional neural network architecture for speech commands recognition,''
  \emph{Proc. Interspeech 2020}, 2020.

\bibitem[He et~al.(2016)He, Zhang, Ren, and Sun]{he2015}
K.~He, X.~Zhang, S.~Ren, and J.~Sun, ``Deep residual learning for image
  recognition,'' in \emph{Proceedings of the IEEE conference on computer vision
  and pattern recognition}, 2016, pp. 770--778.

\bibitem[Chollet(2017)]{chollet2017xception}
F.~Chollet, ``Xception: Deep learning with depthwise separable convolutions,''
  in \emph{CVPR}, 2017.

\bibitem[Kaiser et~al.(2017)Kaiser, Gomez, and Chollet]{kaiser2017depthwise}
L.~Kaiser, A.~Gomez, and F.~Chollet, ``Depthwise separable convolutions for
  neural machine translation,'' \emph{arXiv:1706.03059}, 2017.

\bibitem[Ding et~al.(2020)Ding, Wang, Chang, Wan, and Moreno]{ding2020personal}
S.~Ding, Q.~Wang, S.~Chang, L.~Wan, and I.~Moreno, ``Personal {VAD}:
  Speaker-conditioned voice activity detection,'' \emph{arXiv:1908.04284},
  2020.

\bibitem[Warden(2018)]{warden2018speech}
P.~Warden, ``Speech commands: A dataset for limited-vocabulary speech
  recognition,'' \emph{arXiv:1804.03209}, 2018.

\bibitem[Font et~al.(2013)Font, Roma, and Serra]{freesound}
F.~Font, G.~Roma, and X.~Serra, ``Freesound technical demo,'' in \emph{ACM
  International Conference on Multimedia (MM{\textquoteright}13)}, 2013.

\bibitem[{Park} et~al.(2019){Park}, {Chan}, {Zhang}, {Chiu}, {Zoph}, {Cubuk},
  and {Le}]{park2019}
D.~{Park}, W.~{Chan}, Y.~{Zhang}, C.~{Chiu}, B.~{Zoph}, E.~{Cubuk}, and
  Q.~{Le}, ``{SpecAugment}: A simple data augmentation method for automatic
  speech recognition,'' \emph{arXiv:1904.08779}, 2019.

\bibitem[DeVries and Taylor(2017)]{devries2017specutout}
T.~DeVries and G.~Taylor, ``Improved regularization of convolutional neural
  networks with cutout,'' \emph{arXiv:1708.04552}, 2017.

\bibitem[Sutskever et~al.(2013)Sutskever, Martens, Dahl, and Hinton]{momentum}
I.~Sutskever, J.~Martens, G.~Dahl, and G.~Hinton, ``On the importance of
  initialization and momentum in deep learning,'' in \emph{ICML}, 2013.

\bibitem[He et~al.(2019)He, Zhang, Zhang, Zhang, Xie, and Li]{he2019bag}
T.~He, Z.~Zhang, H.~Zhang, Z.~Zhang, J.~Xie, and M.~Li, ``Bag of tricks for
  image classification with convolutional neural networks,'' in \emph{CVPR},
  2019.

\bibitem[Kuchaiev et~al.(2019)Kuchaiev, Li, Nguyen, Hrinchuk, Leary, Ginsburg,
  Kriman, Beliaev, Lavrukhin, Cook, Castonguay, Popova, Huang, and
  Cohen]{nemo2019}
O.~Kuchaiev, J.~Li, H.~Nguyen, O.~Hrinchuk, R.~Leary, B.~Ginsburg, S.~Kriman,
  S.~Beliaev, V.~Lavrukhin, J.~Cook, P.~Castonguay, M.~Popova, J.~Huang, and
  J.~Cohen, ``{NeMo}: a toolkit for building ai applications using neural
  modules,'' \emph{arXiv:1909.09577}, 2019.

\bibitem[Kingma and Ba(2014)]{kingma2014adam}
D.~P. Kingma and J.~Ba, ``Adam: A method for stochastic optimization,''
  \emph{arXiv:1412.6980}, 2014.

\bibitem[Sehgal and Kehtarnavaz(2018)]{smartapp}
A.~Sehgal and N.~Kehtarnavaz, ``A convolutional neural network smartphone app
  for real-time voice activity detection,'' \emph{IEEE Access}, vol.~6, pp.
  9017--9026, 2018.

\end{thebibliography}
\end{document}